# Invariant Correlations in Simplicial Gravity

*Herbert W. Hamber* [1]

Department of Physics
University of California at Irvine
Irvine, Ca 92717 USA

**ABSTRACT**

Some first results are presented regarding the behavior of invariant correlations in simplicial gravity, with an action containing both a bare cosmological term and a lattice higher derivative term. The determination of invariant correlations as a function of geodesic distance by numerical methods is a difficult task, since the geodesic distance between any two points is a function of the fluctuating background geometry, and correlation effects become rather small for large distances. Still, a strikingly different behavior is found for the volume and curvature correlation functions. While the first one is found to be negative definite at large geodesic distances, the second one is always positive for large distances. For both correlations the results are consistent in the smooth phase with an exponential decay, turning into a power law close to the critical point at $G_c$. Such a behavior is not completely unexpected, if the model is to reproduce the classical Einstein theory at distances much larger than the ultraviolet cutoff scale.

---

[1] Address after January 1,1994: Theory Division, CERN, CH-1211 Genève 23, Switzerland.

# 1  Introduction

Regge's formulation of gravity is the natural discretization for general relativity [1]. At the classical level, it is the only model known to reproduce in four dimensions general relativity as we know it, with continuous curvatures, classical gravitational waves, and no graviton doubling problem in the weak field limit. The correspondence with continuum gravity is particularly transparent in the lattice weak field expansion, with the invariant edge lengths playing the role of infinitesimal geodesics in the continuum. In the limit of smooth manifolds with small curvatures, the continuous diffeomorphism invariance of the continuum theory is recovered [2, 3]. But in contrast to ordinary lattice gauge theories, the model is formulated entirely in terms of coordinate invariant quantities, the edge lengths, which form the elementary degrees of freedom in the theory [4, 2].

Recent work based on Regge's simplicial formulation of gravity has shown in pure gravity the appearance in four dimensions of a phase transition in the bare Newton's constant, separating a *smooth phase* with small negative average curvature from a *rough phase* with large positive curvature [5, 6]. While the fractal dimension seems rather small in the rough phase, indicating a tree-like geometry for the ground state, it is very close to four in the smooth phase close to the critical point. A calculation of the critical exponents in the smooth phase and close to the critical point seems to suggest that the transition is second order, at least for sufficiently large higher derivative coupling, with divergent curvature fluctuations, and that a lattice continuum might therefore be constructed.

In this paper we will present some first result regarding the nature of invariant correlations in simplicial gravity, as derived from numerical studies on lattices with $24 \times 8^4 = 98,304$ and $24 \times 16^4 = 1,572,864$ simplices. The paper is organized as follows. In Sec. 2 we introduce the simplicial action and measure for the gravitational degrees of freedom. We then discuss in Sec. 3 the formulation and expected properties of invariant gravitational correlations, such as the volume-volume and curvature-curvature correlation functions at fixed geodesic distance. In Sec. 4 we present our results and their interpretation, and Sec. 5 then contains our conclusions.

## 2  Action and Measure

We write the four-dimensional pure gravity action on the lattice as

$$I_g[l] = \sum_{\text{hinges } h} V_h \left[ \lambda - k\, A_h \delta_h / V_h + a\, A_h^2 \delta_h^2 / V_h^2 \right], \qquad (2.1)$$

where $V_h$ is the volume per hinge (which is represented by a triangle in four dimensions), $A_h$ is the area of the hinge and $\delta_h$ the corresponding deficit angle, proportional to the curvature at $h$. All geometric quantities can be evaluated in terms of the lattice edge lengths $l_{ij}$, which uniquely specify the lattice geometry for a fixed incidence matrix. The underlying lattice structure is chosen to be hypercubic, with a natural simplicial subdivision to ensure its overall rigidity [2, 7, 8, 9]. In the classical continuum limit the above action is then equivalent to

$$I_g[g] = \int d^4x \sqrt{g} \left[ \lambda - \tfrac{1}{2} k\, R + \tfrac{1}{4} a\, R_{\mu\nu\rho\sigma} R^{\mu\nu\rho\sigma} + \cdots \right], \qquad (2.2)$$

with a bare cosmological constant term (proportional to $\lambda$), the Einstein-Hilbert term ($k = 1/8\pi G$), and a higher derivative term proportional to $a$. For an appropriate choice of bare couplings, the above lattice action is bounded below, due to the presence of the higher derivative term. When considering small fluctuations about a regular lattice the unboundedness present for $a = 0$ is slightly ameliorated by the presence of a lattice momentum cutoff, which cuts off the conformal mode fluctuations at high momenta [2]. Away from almost regular lattices the situation is more complex and one has to resort to non-perturbative methods to investigate the stability of the ground state. For non-singular measures and in the presence of the $\lambda$-term a stable lattice can be shown to arise naturally [8, 7, 9]. The higher derivative terms can be set to zero ($a = 0$), but they nevertheless seem to be necessary for reaching the lattice continuum limit, and are in any case generated by radiative corrections already in weak coupling perturbation theory.

The purely gravitational measure contains an integration over the elementary lattice degrees of freedom, the edge lengths. For the edges we write the lattice integration measure as [4, 7, 8, 9]

$$\int d\mu[l] = \prod_{\text{edges } ij} \int_0^\infty V_{ij}^{2\sigma}\, dl_{ij}^2\, F[l] \;, \qquad (2.3)$$

where $V_{ij}$ is the 'volume per edge', $F[l]$ is a function of the edge lengths which enforces the higher-dimensional analogs of the triangle inequalities, and the power $\sigma = 0$ for the lattice analog of the DeWitt measure for pure gravity. The factor $V_{ij}^{2\sigma}$ plays a role analogous the factor $(\sqrt{g})^{2\sigma}$ which appears for continuum measures [10, 11]. A variety of measures have been proposed in the continuum [10, 11, 12, 13] and on the lattice [14], some of which are even non-local. Since there is no exact gauge invariance on the Regge lattice (nor in any other known lattice formulation of gravity), one cannot uniquely decide a priori which is the most appropiate gravitational measure. Note that *no* cutoff is imposed on small or large edge lengths, if a non-singular measure such as $dl^2$ is used. This fact is essential for the recovery of diffeomorphism invariance close to the critical point, where on large lattices a few rather long edges, as well as some rather short ones, start to appear [6]. The influence of the measure and the dependence of the results on the underlying lattice structure have also been investigated recently in [15].

## 3   Invariant Correlations

Previous work dealt almost exclusively with averages of invariant local operators such as the volume, the curvature and their fluctuations. A great deal of information about the theory can be obtained by considering just these local quantities [6]. But in general the information obtained is restricted only to the leading long distance properties, and higher order corrections as well as additional information can only be obtained by considering correlations between operators separated by some geodesic distance [8, 6]. In quantized gravity complications arise due to the fact that the physical distance between any two points $x$ and $y$ in a fixed background geometry,

$$d(x,y\,|\,g) \;=\; \min_\xi \int_{\tau(x)}^{\tau(y)} d\tau \sqrt{g_{\mu\nu}(\xi)\tfrac{d\xi^\mu}{d\tau}\tfrac{d\xi^\nu}{d\tau}} \;, \qquad (3.1)$$

is a fluctuating quantity dependent on the background metric. In addition, the Lorentz group used to classify spin states is meaningful only as a local concept. Since the simplicial formulation is completely coordinate independent, the introduction of the local Lorentz group requires the definition of a tetrad within each simplex, and the notion of a spin connection to describe the parallel transport of tensors between flat simplices. Some of these aspects have also been discussed recently in the continuum in [16, 17].

For a set of local operators $\{\tilde{O}_\alpha(x)\}$, we can consider in general the set of connected correlations at fixed geodesic distance $d$,

$$G_{\alpha\beta}(d) \;=\; <\tilde{O}_\alpha(x)\,\tilde{O}_\beta(y)\,\delta(|x-y|-d)>_c \;. \qquad (3.2)$$

A suitable set of local operators in the continuum is represented, for example, by the fourteen algebraically independent coordinate scalars which can be constructed from the components of the Riemann tensor

$$R(x), \ R_{\mu\nu\lambda\sigma} R^{\mu\nu\lambda\sigma}(x), \ R_{\mu\nu} R^{\mu\nu}(x), \ R^2(x), \ \dots \tag{3.3}$$

On the lattice one can construct discrete approximations to these operators [7]. Since the deficit angles are proportional to the Gaussian curvatures associated with the hinges, in general more than one hinge needs to be considered, and the corresponding lattice operators are not completely local, in the sense that they can involve a number of neighboring hinges, as well as the angles describing their relative orientation. For a small space-time loop we may make the identification

$$R_{\mu\nu\rho\sigma} \, \Sigma^{\rho\sigma} \ \sim \ \delta_h \, U^{(h)}_{\mu\nu} \tag{3.4}$$

where $\Sigma^{\rho\sigma}$ is the area bivector of the loop, and $U^{(h)}_{\mu\nu}$ is a bivector perpendicular to the hinge $h$,

$$U^{(h)}_{\mu\nu} = \frac{1}{2A_h} \, \epsilon_{\mu\nu\rho\sigma} \, l^\rho_{(a)} l^\sigma_{(b)} \tag{3.5}$$

with $l^\rho_{(a)}$ and $l^\rho_{(b)}$ the two vectors forming two sides of the hinge $h$. This relation emphasizes the fact that the deficit angle gives only information about the projection of the Riemann tensor in the plane of the (small) loop $C$ orthogonal to the hinge.

On the lattice the natural choices for invariant operators are

$$\begin{aligned}
\sqrt{g}\,(x) \ &\to \ \sum_{\text{hinges}\, h \supset x} V_h \\
\sqrt{g}\, R(x) \ &\to \ 2 \sum_{\text{hinges}\, h \supset x} \delta_h A_h \\
\sqrt{g}\, R_{\mu\nu\lambda\sigma} R^{\mu\nu\lambda\sigma}(x) \ &\to \ 4 \sum_{\text{hinges}\, h \supset x} (\delta_h A_h)^2 / V_h
\end{aligned} \tag{3.6}$$

(we have omitted here on the r.h.s. an overall numeric coefficient, which will depend on how many hinges are actually included in the summation; if the sum extends over all hinges within a single hypercube, then there will be a total of 50 hinge contributions). From these formulae, representations for operators such as $R^2$ can then be constructed,

$$\sqrt{g}\, R^2(x) \ \to \ \left( \sum_{\text{hinges}\, h' \supset x} V_{h'} \right) \left( \sum_{\text{hinges}\, h \supset x} \delta_h A_h / V_h \right)^2 . \tag{3.7}$$

If the deficit angles are averaged over a number of contiguous hinges which share a common vertex, one is naturally lead to the correlation

$$G_R(d) \ \sim \ < \sum_{h \supset x} \delta_h A_h \sum_{h' \supset y} \delta_{h'} A_{h'} \, \delta(|x-y|-d) >_c , \tag{3.8}$$

which corresponds to correlations in the scalar curvature,

$$\sim < \sqrt{g}\, R(x)\, \sqrt{g}\, R(y)\, \delta(|x-y|-d) >_c . \qquad (3.9)$$

Similarly one can construct

$$G_V(d) \sim < \sum_{h \supset x} V_h \sum_{h' \supset y} V_{h'}\, \delta(|x-y|-d) >_c , \qquad (3.10)$$

which corresponds to the volume correlations

$$\sim < \sqrt{g}(x)\, \sqrt{g}(y)\, \delta(|x-y|-d) >_c . \qquad (3.11)$$

A more precise definition of what is meant by the above correlation functions on the lattice is as follows. First, for a given operator, the geodesic distance between any two points $x$ and $y$ is determined in a fixed background geometry (the actual method to do this will be described in detail below). Next the correlations are computed for all pairs of points within geodesic distance $d$ and $d + \Delta d$, where $\Delta d$ is an interval comparable or slightly larger than the average lattice spacing $l_0 = \sqrt{<l^2>}$, but clearly much smaller than the distance $d$ considered. Finally, the correlations determined for a fixed geodesic distance $d$ are averaged over all the metric configurations, and the connected parts are then computed in the usual way.

In general the above correlations will contain particles of different spin (0,2), but at large distances the lightest state with spin two (the graviton) should dominate, if the theory is to reproduce classical general relativity at large distances.

Up to now we have considered correlations between scalar densities. For higher rank tensor densities things become more complicated. In general one can compare two vectors at different locations only if one of them is first parallel transported to the other's location. Integrating the parallel transport equation for an arbitrary four-vector $S_\lambda$,

$$\frac{dS_\mu}{d\tau} = \Gamma^\lambda_{\mu\nu} \frac{dx^\nu}{d\tau} S_\lambda , \qquad (3.12)$$

along some path $C$ connecting $x$ and $y$, one obtains

$$S_\mu(y) = [P \exp \int_x^y \Gamma_{\cdot\lambda}(z) dz^\lambda]_\mu^{\ \nu}\, S_\nu(x) . \qquad (3.13)$$

This path-dependent factor then enters in the definition of the correlation functions, for example involving two coordinate vectors $U$ and $V$,

$$G_{UV}(d) = < U^\mu(x)\, [P \exp \int_x^y \Gamma_{\cdot\lambda}(z) dz^\lambda]_\mu^{\ \nu}\, V_\nu(y)\, \delta(|x-y|-d) >_c . \qquad (3.14)$$

It is clear that such a correlation function will in general be path-dependent. In addition, one has the problem that the Regge formulation is coordinate independent, and frames have to be introduced within each simplex to specify the orientation of a vector, and in order to define an affine connection. If we consider a single closed path $C$, we obtain the analog of the Wilson loop

$$W[C] \equiv \, < \text{Tr}[P \exp \oint_C \Gamma^{\cdot}_{\cdot \nu} dx^\nu \, - \, 1] > \qquad (3.15)$$

It is convenient to consider planar loops, which are spanned by the geodesics tangent to a plane at some point in the center of the loop. On the simplicial lattice the Wilson loop is computed by evaluating the deficit angle, (and its moments), associated with a large planar loop,

$$< (\delta_C)^n > \, = \, < (\sum_{s \subset C} \theta_s - 2\pi)^n > \qquad (3.16)$$

where $s$ labels the simplices traversed by the loop, and $\theta_s$ is the appropriate internal dihedral angle. Like the deficit angle $\delta_h$ itself, this quantity is of course coordinate independent (it describes the rotation angle of a vector which is parallel-transported around the loop). In gravity the Wilson loop does not have quite the same interpretation as in ordinary gauge theories, since it is not directly associated with the Newtonian potential energy of two static bodies [17, 6]. In ordinary gauge theories at strong coupling the Wilson loop decays like the area of the loop, due to the strong independent fluctuations of the gauge fields at different points in space-time and ensuing cancellations. In gravity the situation is quite different, since the connections cannot be considered as independent variables, and the fluctuations in the deficit angles at different points in space-time are strongly correlated. In the remainder of this paper we shall restrict our attention to correlations between local operators that transforms like coordinate scalars, such as the ones described at the beginning of this section, and leave the study of the other, more complicated, correlation functions for future work.

It is well known that in the weak-field expansion the Einstein-Hilbert action contains both spin two (graviton) and spin zero (conformal mode) contributions. By a judicious choice of invariant correlation functions, one would like to be able to isolate the physical properties of the graviton and the conformal mode. The same result holds of course on the lattice, as can be seen by expanding the Regge action about a regular lattice, since the lattice and continuum actions are equivalent for sufficiently smooth manifolds [2, 3]. In the continuum, after expanding the metric around flat space (which requires $\lambda = 0$),

$$g_{\mu\nu} \, = \, \eta_{\mu\nu} + \sqrt{16\pi G} \, h_{\mu\nu}, \qquad (3.17)$$

one can cast the lowest order, quadratic, contribution to the action in the form

$$I_E[h_{\mu\nu}] = \tfrac{1}{2} \int d^4x \, h_{\mu\nu} V_{\mu\nu\lambda\sigma} h_{\lambda\sigma} \qquad (3.18)$$

where $V$ is a matrix which can be expressed in terms of spin projection operators. In momentum space it can be written as

$$V = \left[ P^{(2)} - 2 P^{(0)} \right] p^2 \qquad (3.19)$$

where $P^{(2)}$ and $P^{(0)}$ are spin two and spin zero projection operators introduced in [18]. Physically, the two terms correspond to the propagation of the graviton and of the conformal mode, respectively, with the latter one appearing with the 'wrong' sign. In the 'Landau' gauge, with a gauge fixing term $\alpha^{-1}(\partial_\mu h^{\mu\nu})^2$ and $\alpha = 0$, one obtains for the graviton propagator in momentum space

$$G_{\mu\nu\lambda\sigma}(p) = \frac{P^{(2)}_{\mu\nu\lambda\sigma}}{p^2} - \frac{\frac{1}{2} P^{(0)}_{\mu\nu\lambda\sigma}}{p^2} \qquad (3.20)$$

The unboundedness of the Euclidean gravitational action shows up clearly in the weak field expansion, since the spin zero mode acquires a kinetic or propagator term with the wrong sign. We note here also the fact that the conformal mode is not pure gauge. In the presence of higher derivative terms in the gravitational action, the above result is modified by terms $O(p^4)$, and becomes [19]

$$G_{\mu\nu\lambda\sigma}(p) = \frac{P^{(2)}_{\mu\nu\lambda\sigma}}{p^2 + \frac{2a}{k} p^4} + \frac{\frac{1}{2} P^{(0)}_{\mu\nu\lambda\sigma}}{-p^2 + \frac{a}{k} p^4} \qquad (3.21)$$

On the lattice, all terms in Eq. (2.1) contain higher derivative contribution; in the last term proportional to $a$ these are the leading contribution. Of course the appearance of higher derivative terms is unavoidable in lattice field theories [20]. It was shown in [2, 3] that the simplicial lattice model, when expanded in the weak field limit about a regular lattice with hypercubic connectivity, gives rise at long wavelengths to the same graviton propagator as the continuum theory. We should add that in the presence of a cosmolgical constant, flat space is no longer a solution of the classical equations of motion, and the above expansion for the metric looses part of its meaning due to the presence of the tadpole term (the appearence of a 'graviton mass' in this case appears as an artifact of the expansion about the wrong vacuum).

A priori one cannot exclude to possibility that some of the states acquire a mass away from the critical point. In the presence of massive excitations one would expect the following behavior for the correlation functions of Eq. 3.2,

$$G_{\alpha\beta}(d) \underset{d \gg 1/m}{\sim} T_{\alpha\beta} \, d^{-\sigma} \exp(-md) + T'_{\alpha\beta} \, d^{-\sigma'} \exp(-m'd) + \cdots \qquad (3.22)$$

where $m$ and $m'$ are the masses associated with the lowest lying excitations (here with $m < m'$). Close to the critical point in Newton's constant at $k = k_c$, where one expects the graviton mass to vanish, the leading behavior should turn into a power law

$$G_{\alpha\beta}(d) \underset{d \gg l_0}{\sim} T_{\alpha\beta} \frac{C}{d^\sigma} \qquad (3.23)$$

where $T$ is some numeric matrix with entries of order one, $C$ is a constant that sets the overall scale, $\sigma$ a power which depends on the scaling dimensions of the chosen operators, and $l_0$ the average lattice spacing. Furthermore on physical grounds one would expect, based on the structure of the graviton propagator which emerges from the weak field expansion, that correlations in the volume should behave rather differently from correlations in the curvature. This is indeed what was found for the zero momentum components, the volume and curvature fluctuations discussed in [6].

Some partial information about the behavior of the relevant correlations can be obtained indirectly from local averages. In [5] gravitational physical observables, such as the average curvature and its fluctuation, were introduced. The appropriate lattice analogs of these quantities are readily written down by making use of the usual correspondences $\int d^4x \sqrt{g} \to \sum_{\text{hinges } h} V_h$ etc.. On the lattice one prefers to define quantities in such a way that variations in the average lattice spacing $\sqrt{<l^2>}$ are compensated by the appropriate factor as determined from dimensional considerations. In the case of the average curvature we have defined the lattice quantity $\mathcal{R}$ as

$$\mathcal{R}(\lambda, k, a) = <l^2> \frac{<2\sum_h \delta_h A_h>}{<\sum_h V_h>} \sim \frac{<\int \sqrt{g}\, R>}{<\int \sqrt{g}>}, \qquad (3.24)$$

and similarly for the curvature fluctuation,

$$\chi_{\mathcal{R}}(\lambda, k, a) \sim \frac{<(\int \sqrt{g}\, R)^2> - <\int \sqrt{g}\, R>^2}{<\int \sqrt{g}>} \sim \frac{1}{V} \frac{\partial^2}{\partial k^2} \ln Z . \qquad (3.25)$$

The curvature critical exponent $\delta$ is introduced via

$$\mathcal{R} \underset{k \to k_c}{\sim} -A_{\mathcal{R}} (k_c - k)^\delta \qquad (3.26)$$

Then it is easy to see that the curvature fluctuation is related to the connected scalar curvature correlator at zero momentum

$$\chi_{\mathcal{R}} \sim \frac{\int d^4x \int d^4y <\sqrt{g}R(x)\sqrt{g}R(y)>_c}{<\int d^4x \sqrt{g}>} \underset{k \to k_c}{\sim} (k_c - k)^{\delta-1} . \qquad (3.27)$$

A divergence in the fluctuation is then indicative of long range correlations, corresponding to the presence of a massless particle. Close to the critical point one would expect for large separations a power law decay in the geodesic distance,

$$<\sqrt{g}R(x)\sqrt{g}R(y)> \underset{|x-y| \to \infty}{\sim} \frac{1}{|x-y|^{2n}}, \qquad (3.28)$$

with the power $n$ related to the exponent $\delta$ via $n = \delta d/(1+\delta) = d - 1/\nu$, and with the exponent $\nu$ defined as $\nu = (1+\delta)/d$.

One can contrast the behavior of the preceding quantities, associated strictly with the curvature, with the analogous quantities involving the local volumes, which correspond to the square root of the determinant of the metric in the continuum. One considers the average volume $<V>$, and its fluctuation defined as

$$\chi_V(\lambda, k, a) \sim \frac{<(\int \sqrt{g})^2> - <\int \sqrt{g}>^2}{<\int \sqrt{g}>} \sim \frac{1}{V} \frac{\partial^2}{\partial \lambda^2} \ln Z. \qquad (3.29)$$

The latter is then related to the connected volume correlator at zero momentum

$$\chi_V \sim \frac{\int d^4x \int d^4y <\sqrt{g(x)}\sqrt{g(y)}>_c}{<\int d^4x \sqrt{g}>}. \qquad (3.30)$$

It is clear that fluctuations in the curvature are sensitive to the presence of a spin two massless particle, while fluctuations in the volume probe only the correlations in the scalar, conformal mode channel. From the previous discussion it should be evident that in the case of gravity a significant qualitative difference should be expected in the two types of correlations, given in particular the 'wrong' sign for the kinetic term associated with the conformal mode in perturbation theory. As will be shown below, such a distinctive behavior is indeed seen.

To conclude this section, let us discuss briefly the method of computing the geodesic distance between two given points in a fixed background metric configuration. We can write schematically the propagator for the scalar field in a *fixed* background geometry, specified by some distribution of edge lengths, as

$$G(d) = <y| \frac{1}{-\partial^2 + m^2} |x>, \qquad (3.31)$$

where $d$ is the geodesic distance between the two spacetime points $x, y$, and $\partial^2$ is the usual covariant Laplacian,

$$\partial^2 \phi \equiv \frac{1}{\sqrt{g}} \partial_\mu \sqrt{g}\, g^{\mu\nu} \partial_\nu \phi. \qquad (3.32)$$

Now fix one point at the origin 0, and use the discretized form of the scalar field action used in Ref. [21],

$$I_\phi[l, \phi] = \tfrac{1}{2} \sum_{<ij>} V_{ij} \left(\frac{\phi_i - \phi_j}{l_{ij}}\right)^2 + \tfrac{1}{2} m^2 \sum_i V_i \phi_i^2, \qquad (3.33)$$

Here $V_{ij}$ is the baricentric volume associated with the edge $l_{ij}$, while $V_i$ is a baricentric volume associated with the site $i$. Then the discrete equation of motion for the field $\phi_i$ in the presence of a $\delta$-function source of unit strength localized at the origin gives us the sought-after Green's function. It is useful to write this equation in a form suitable for iteration, by solving explicitly for $\phi_i$,

$$\phi_i = \frac{1}{W_i}\Big(\sum_{j \neq i} W_{ij}\, \phi_j + \delta_{i0}\Big), \tag{3.34}$$

with weights given by

$$W_i = \sum_{j \neq i} \Big(\frac{m^2}{2} + \frac{1}{l_{ij}^2}\Big) V_{ij} \ , \qquad W_{ij} = \frac{V_{ij}}{l_{ij}^2}. \tag{3.35}$$

Here the sums extend over nearest-neighbor points only, and $\delta_{i0}$ is a delta-function source localized at the origin on site 0. The above equation for $\phi_i$ can then be solved by an iterative procedure, taking for example $\phi_i = 0$ as an initial guess. After the solution $\phi_i$ has been determined by relaxation, at large distances from the origin one has

$$\phi_i \sim G(d(i,0)) \underset{d(i,0) \gg 1/m}{\sim} A\sqrt{m/d(i,0)^3}\, \exp(-md(i,0)), \tag{3.36}$$

which determines the geodesic distance $d(i,0)$ from lattice point 0 to lattice point $i$. This method is far more efficient and accurate than trying to determine the geodesic distance by sampling paths connecting the two points, as was done in [6], but is equivalent to it at large distances [22].

Still an important correction needs to be applied. In general the asymptotic behavior of the above scalar correlation is reached only for large distances compared to the quantity $1/m$. At relatively short distances, comparable to the average spacing between lattice sites, sizable deviations appear and corrections have to be applied. Define a 'propagator distance' $d_{pr}$ between two points labeled by 0 and $i$ as

$$d_{pr}(i,0) = -\frac{1}{m}\log\Big(\frac{G_i}{G_0}\Big) \tag{3.37}$$

where $m$ is taken to be of order one (the specific value is not important here). $d_{pr}$ then becomes proportional to the true geodesic distance only at large distances. Next call $d_{rw}$ the 'random walk' or true geodesic distance between the two points, obtained by adding up the lengths of the line segments forming the shortest lattice path from 0 to $i$. As can be seen clearly from Fig. 1, in spite of some transients for small distances, the two quantities are rather smooth functions of each other, and quite independently of the binning procedure used. For the purpose of this work, a quadratic polynomial interpolation of the true physical distance $d_{rw}$ in terms of the quantity $d_{pr}$ is quite adequate, bringing down the uncertainty in the distance determination significantly below the statistical fluctuations in the correlations. We have checked that this correction is, as expected, almost independent of the geometry configuration, for $k < k_c$ and fixed values of $\lambda$ and $a$. The above method allows one to estimate the geodesic distance reasonably accurately even for points which are quite far apart, and for which the determination of the true physical distance by the shortest path method would be prohibitively time consuming. In a number of cases we have explicitly checked that indeed the correlation functions at not too short distances become insensitive within errors to the method used for determining the distance. A detailed comparison of correlation functions obtained by the two different methods will be presented elsewhere. But for large distances it is clear that the propagator method is far more efficient.

## 4 Results

As in our previous work, the edge lengths are updated by a straightforward Monte Carlo algorithm, generating eventually an ensemble of configurations distributed according to the action of Eq. (2.1) and measure of Eq. (2.3). Further details of the method as applied to pure gravity are discussed in [23], and will not be repeated here. In this work the edge length configurations already generated in [6] were used as a starting point.

For computing the correlation functions, we considered lattices of size $8 \times 8 \times 8 \times 8$ (with 4096 sites, 61440 edges, 98304 simplices) and $16 \times 16 \times 16 \times 16$ (with 65536 sites, 983040 edges, 1572864 simplices). Even though these lattices are not very large, one should keep in mind that due to the simplicial nature of the lattice there are many edges per hypercube with many interaction terms, and as a consequence the statistical fluctuations can be comparatively small, unless measurements are taken very close to a critical point, and at rather large distances in the case of the correlations. The usefulness of studying two different lattice sizes lies in the fact that finite size effects can be systematically monitored.

As usual the topology was restricted to a four-torus (periodic boundary conditions). We have argued before that one could perform similar calculations with lattices employing different boundary conditions or topology, but the universal infrared scaling properties of the theory should be determined only by short-distance renormalization effects. The renormalization group equations are in fact *independent* of the boundary conditions, which enter only in their solution as it affects the correlation functions through the presence of a new dimensionful parameter, the linear system size $L = V^{1/4}$.

In this work the bare cosmological constant $\lambda$ appearing in the gravitational action of Eq. (2.1) was fixed at 1 (this coupling sets the overall scale in the problem). The higher derivative coupling $a$ was set to 0 and 0.005, in order to investigate the sensitivity of the results to what is expected to be an irrelevant term, at least for small $a$. For the measure of Eq. (2.3) this choice of parameters leads to a well behaved ground state for $k < k_c \approx 0.063$ (for $a = 0$) or $k < k_c \approx 0.245$ (for $a = 0.005$). The system then resides in the 'smooth' phase, with a fractal dimension close to four; on the other hand for $k > k_c$ the curvature becomes very large ('rough' phase), and the lattice tends to collapse into degenerate configurations with very long, elongated simplices. For $a = 0$ we investigated six values of $k$ $(0.00, 0.01, 0.02, 0.03, 0.04, 0.05)$, while for $a = 0.005$ we looked at five values of $k$ $(0.00, 0.05, 0.10, 0.15, 0.20)$.

From physical considerations it seems reasonable to impose the constraint that the scale of the curvature in magnitude should be much smaller than the average lattice spacing, but much larger than the size of the system, or in other words

$$< l^2 > \ll \; < l^2 > |\mathcal{R}|^{-1} \ll V^{1/2}. \tag{4.1}$$

This is equivalent to the statement that in momentum space the physical scales should be much smaller that the ultraviolet cutoff, but much larger than the infrared one. This fact prevent us from studying larger values of $a$, since the curvature then becomes too small. Conversely, for too small values of $a$ (and in particular negative $a$) the curvature becomes rather large in magnitude, and the results become useless when one is too far away from the critical point in $k$. The above constraint then requires that either $k$ is very close to $k_c$, or that $a$ cannot be too small.

For both values of the coupling $a$ we studied correlations on both a $8^4$ and a $16^4$ lattice. On the $8^4$ lattice we generated 900-1600 consecutive configurations at $a = 0$ and 200-400 configurations at $a = 0.005$, for each value of $k$. On each configuration all correlations were measured for 20 origins, chosen at random. On the $16^4$ lattice we generated 180 configurations at $a = 0$ and about 300 configurations at $a = 0.005$, for each value of $k$. Results for a larger statistical sample are in progress and will be presented elsewhere. On each configuration all correlations were measured for 256 origins, again chosen at random. The runs are comparatively longer on the larger lattices, since it was possible in that case to use a fully parallel version of the program. All the geodesic distance and correlation measurements were also done in parallel. The propagator distances were computed with $m = 1$ and 40 relaxation iterations, which proved completely adequate for our purposes. As a check, we computed in a number of instances the correlations using the 'random walk' or exact lattice geodesic distance, and found no discrepancy within errors for distance larger than one or two average lattice spacings. The correlation functions were further divided in bins according to the value of the geodesic distance, with a bin width $\Delta d = 2$, which is comparable to the average lattice spacing (which is about $l_0 \equiv \sqrt{<l^2>} \approx 2.3$ for the couplings used in this work, almost independently of $a$ and $k$).

We computed the correlations for all the operators listed in Eqs. (3.6) and (3.7). In constructing these operators, the summation over hinges was restricted to all hinges $h$ which are within one hypercube containing a given site labeled by $x$ (there are 50 such hinges). Here we will only present the results for the volume-volume correlation $G_V(d)$ and the curvature-curvature correlation $G_R(d)$; the other correlations are more difficult to determine accurately, and the results for those will be presented elsewhere. Not unexpectedly, the correlation functions are more difficult to determine at large distances, where they become small and tend to be drowned in the statistical noise. For $a = 0$ and $k = 0.05$ we did not achieve enough statistics to obtain accurate data for the correlations, and therefore results for this coupling will not be presented here. The results we have obtained so far are shown in Figs. 2 to 5.

The first rather striking result is that the *volume correlations* (shown in Fig. 2 for $a = 0$ and in Fig. 3 for $a = 0.005$) are negative at large distances. This seems to happen for all values of $a$ and $k$ we have investigated. At very short distances, comparable to one or two average lattice spacings, we expect the correlation functions to show some oscillations due to the underlying lattice structure, and this is indeed what is observed. In particular, for $a = 0.005$ the lattice action contains second neighbor interactions, and it makes little sense in this case to analyze correlations which involve distances comparable to two average lattice spacings or less. Indeed some oscillations are observed at relatively small distances both in the volume and curvature correlations, and some oscillations persist also when the geodesic distance is computed by the random-walk method. Of course the oscillations can be reduced and even eliminated by using a larger bin width for the geodesic distance, as explained in the previous section, but then only few points would be left to display (as we said, we have chosen a bin width $\Delta d = 2$, which is comparable to the average lattice spacing, $l_0 \sim 2.3$; on larger lattices a larger bin width could be used).

The *curvature correlations* (shown in Fig. 4 for $a = 0$ and in Fig. 5 for $a = 0.005$) on the other hand are always positive, even though, again, one notices some oscillations at short distances due to the lattice structure. To analyze their behavior further, one can attempt to fit the correlation at 'large' distances, here meaning $d \gg l_0$, to an exponential decay, as indicated in Eq. (3.22). Alternatively, one can try to fit them to a power law close to the critical point at $k_c$, as indicated in Eq. (3.23).

If the correlations are fitted to an exponential decay, one finds that the behavior is always consistent with a mass that decreases as one approaches the critical point. Close to this critical point let us write for the mass of the spin two particle, which is expected to determine the long distance behavior of the curvature-curvature correlation function,

$$\left[m^{(2)}\right]^2 \underset{k \to k_c}{\sim} A^{(2)} (k_c - k) \tag{4.2}$$

and similarly for the spin zero state, which determines the long distance behavior of the volume-volume correlation function,

$$\left[m^{(0)}\right]^2 \underset{k \to k_c}{\sim} A^{(0)} (k_c - k) \tag{4.3}$$

We shall refer to $m^{(0)}$ as the 'mass' of the spin zero state, in spite of the fact that its propagator amplitude has the wrong sign for a real particle. The motivation for using the mass squared in the preceding equation is as follows. In our previous work we estimated the critical exponent $\nu$, which determines how the dynamical graviton mass approaches zero at the critical point, $m \sim (k_c - k)^\nu$, and found that it was close to 1/2 (in Ref. [6] we estimated $\nu \approx 0.4$ for $a = 0.005$). Also it should be added for the sake of clarity, that the values we quote refer to 'physical' masses, and not to masses in units of the lattice spacing, which would be larger by about a factor of two, since, as we mentioned previously, the average lattice is not one but about $l_0 \sim 2.3$.

For both spin zero and spin two, we find a clear decrease in the mass towards the critical point. The decrease is more clearly seen for $a = 0$, since the masses are larger in this case. For the amplitude we estimate $A^{(0)} = 2.5(8)$ and $A^{(2)} = 2.8(9)$ at $a = 0$, and $A^{(0)} = 0.09(6)$ and $A^{(2)} = 0.18(7)$ at $a = 0.005$. The rapid decrease in the mass amplitudes $A^{(0)}$ and $A^{(2)}$, which incidentally are not expected to be universal, is caused by a relatively small change in the bare coupling $a$, and is not completely unexpected in light of the asymptotic freedom of higher derivative gravity theories [19]. Here we are making the assumption that both masses will indeed go to zero at the same critical point. To a certain extent this seems to be justified by the fact that the two masses have comparable values close to $k_c$, a fact that is reflected in the amplitudes, which are also comparable for a given value of $a$. The results for the correlations do not allow one yet to determine in a clean way if this is indeed what is happening, but they are certainly consistent with such an expectation. We will leave a more accurate determination of the mass parameters for future work.

When the mass of the particle is very small, it becomes difficult to distinguish an exponential decay of the correlation function from the pure power behavior of Eq. (3.23), which is expected to appear at $k_c$, at least if coordinate invariance gets fully restored in the neighborhood of the critical point. Close to $k_c$, we have estimated the powers to be $\sigma^{(0)} = 1.4(5)$ and $\sigma^{(2)} = 2.3(8)$ at $a = 0$, and $\sigma^{(0)} = 0.8(5)$ and $\sigma^{(2)} = 2.6(8)$ at $a = 0.005$. While the errors are quite large, the results for the two values of $a$ do not seem to be inconsistent. We leave a more accurate quantitative determination of these numbers for future work. Close to the critical point, the power associated with the volume correlation function is not very far from 2, the free field value (it seems a bit smaller, but this could be due to the smallness of our lattice which makes it difficult to reach the asymptotic behavior of the correlations), while the power associated with the curvature correlation function seems somewhat larger (the estimates for the critical exponent $\delta$ presented in [6] would suggest a value closer to 3, see Eq. (3.28)).

In ref. [6] several values for $a$ were studied, and it was found that while for $a = 0.005$ the results suggest a second order phase transition, for $a = 0$ the singularity in $\mathcal{R}$ appears in fact to be logarithmic ($\delta$ close to 0), suggesting a first order transition with no continuum limit. This situation should reflect itself in a different behavior of the curvature correlation at large distances in the two cases, but the accuracy here is not sufficient to see the difference. Let us also remark here that the lack of a (positive) divergence in the volume fluctuation close to the critical point found in [6] is seen here as a reflection of the fact that the volume correlation is negative at sufficiently large distances. This should give rise to cancellations, and make the singular contribution to the fluctuation eventually negative definite close to the continuum limit.

In conclusion, our results are not inconsistent with the expectation that close to the critical point the long distance properties of the model are described by a theory whose graviton propagator is of the form described in Eq. (3.20), with a vanishing mass for both the spin two and spin zero components. The latter result is of course crucial for the recovery of general coordinate invariance: both contributions must survive in the right proportions if the model is to reduce to classical General Relativity at large distances. A careful study of the above issues should give further support to the argument that coordinate invariance is indeed recovered in this model at large distances, and that the correct low energy theory is recovered.

Let us add that it seems rather remarkable that the results described here seem to hold, in spite of the fact that a) the model includes a bare cosmological constant to start with, b) the model contains lattice higher derivative terms, c) there is no sensible notion of continuous lattice diffeomorphisms away from smooth manifolds, d) there are no classical solutions of Einstein's theory with a $\lambda$-term on a four-torus, and e) the model exists only in one phase (the smooth phase) which appears only for $G > G_c$.

## 5   Conclusions

In the previous sections we have presented some first results regarding the properties of invariant correlations in the context of a model for quantum gravity based on Regge's formulation. The determination of invariant correlations as a function of geodesic distance is a difficult task, since the geodesic distance between any two points is a function of the fluctuating geometry, and is not given a priori. The scalar propagator method provides an efficient method for computing the distances, even though a new source has to be given for each point from which distances are to be determined.

A strikingly different behavior was found for the volume-volume and curvature-curvature correlation functions. The first one is decaying and negative definite at large geodesic distances, while the second one is always positive for large geodesic distances. The behavior found for both correlation functions is consistent with an exponential decay, which eventually turns into a power law close to the critical point at $G_c$. We have argued that such a bizarre behavior is precisely what is expected if the model is supposed to reproduce the classical Einstein theory, for distances which are very large compared to the ultraviolet cutoff scale. It is a consequence of the fact that the weak-field expansion propagator for Einstein's theory, in the absence of a cosmological constant term, contains both spin zero and spin two modes, with the wrong sign for the conformal mode. Our results seem to indicate that, besides the graviton, such a mode will survive with the correct low energy kinetic term. This seems to happen in a model for gravity which at short distances is far removed from the pure Einstein theory, containing both a bare cosmological term and bare higher derivative lattice terms.

**Acknowledgements**

The numerical computations were performed at NCSA and CTC under a *Grand Challenge* allocation grant. The parallel version of the program used in this work was written for the CM5 with Y. Tosa, and his invaluable help is here gratefully acknowledged. This work was supported in part by the NSF under grant PHY-9208386.

# 6   Note added in proof

After this work was submitted for publication, a preprint appeared [24] in which edge and volume correlations are computed in the Regge model as a function of vertex label on lattices of size $3^3 \times 8$ and $4^3 \times 16$. A direct comparison with the results presented in this paper is not possible at this time, since the above authors do not compute correlations at fixed geodesic distance.

# Figure Captions

Fig. 1 Propagator distance $d_{pr}$ versus true physical geodesic distance $d_{rw}$ from the origin, on a typical geometry configuration ($\lambda = 1$, $k = 0.2$, $a = 0.005$). Diamonds represent averages on an $8^4$ lattice, squares represent averages on a $16^4$ lattice. The continuous line is a quadratic interpolation in the region $d_{rw} < 15$. For larger distances, a direct determination of $d_{rw}$ requires a prohibitively large sampling of paths, and finite volume effects become predominant.

Fig. 2 Negative of the volume correlations $G_V(d)$ for $\lambda = 1$ and $a = 0$, and $k = 0.00$ ($\diamond$), 0.01 (+), 0.02 ($\square$), 0.03 ($\times$), and 0.04 ($\triangle$) ($k_c \approx 0.063$). Results are from a $16^4$ lattice, and the lines represent best fits to the data.

Fig. 3 Negative of the volume correlations $G_V(d)$ for $\lambda = 1$ and $a = 0.005$, and $k = 0.00$ ($\diamond$), 0.05 (+), 0.10 ($\square$), 0.15 ($\times$) and 0.20 ($\triangle$) ($k_c \approx 0.245$). Results are from a $16^4$ lattice, and the lines represent best fits to the data.

Fig. 4 Curvature correlations $G_R(d)$ for $\lambda = 1$ and $a = 0$, and $k = 0.00$ ($\diamond$), 0.01 (+), 0.02 ($\square$), 0.03 ($\times$), and 0.04 ($\triangle$) ($k_c \approx 0.063$). Results are from a $16^4$ lattice, and the lines represent best fits to the data.

Fig. 5 Curvature correlations $G_R(d)$ for $\lambda = 1$ and $a = 0.005$, and $k = 0.00$ ($\diamond$), 0.05 (+), 0.10 ($\square$), 0.15 ($\times$) and 0.20 ($\triangle$) ($k_c \approx 0.245$). Results are from a $16^4$ lattice, and the lines represent best fits to the data.

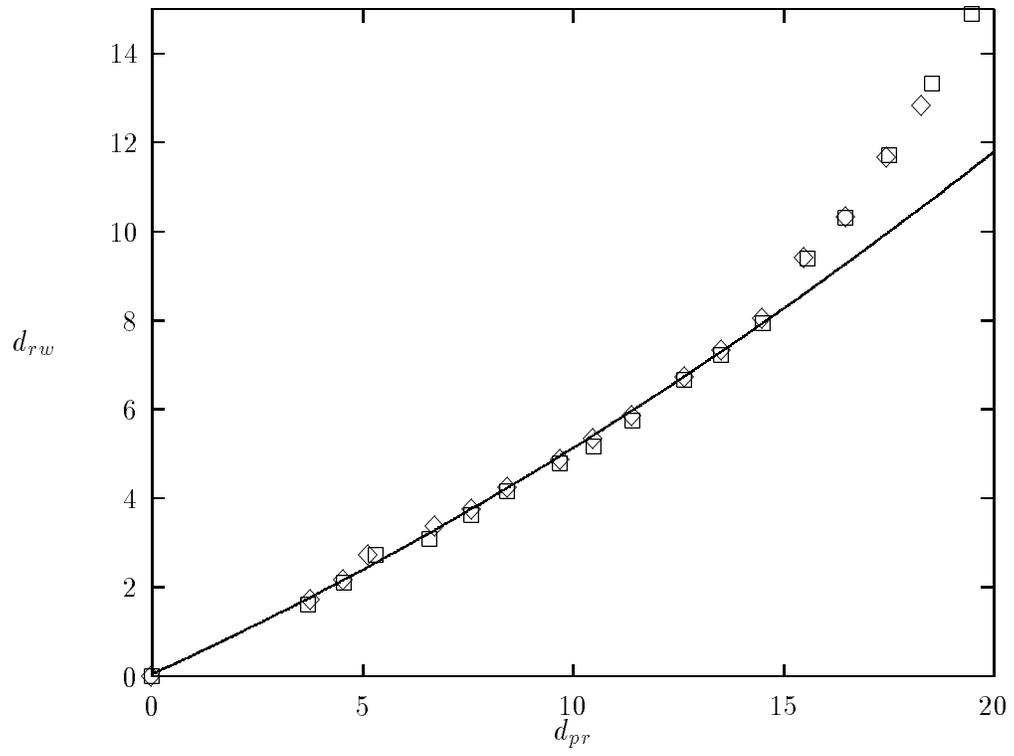

Fig. 1

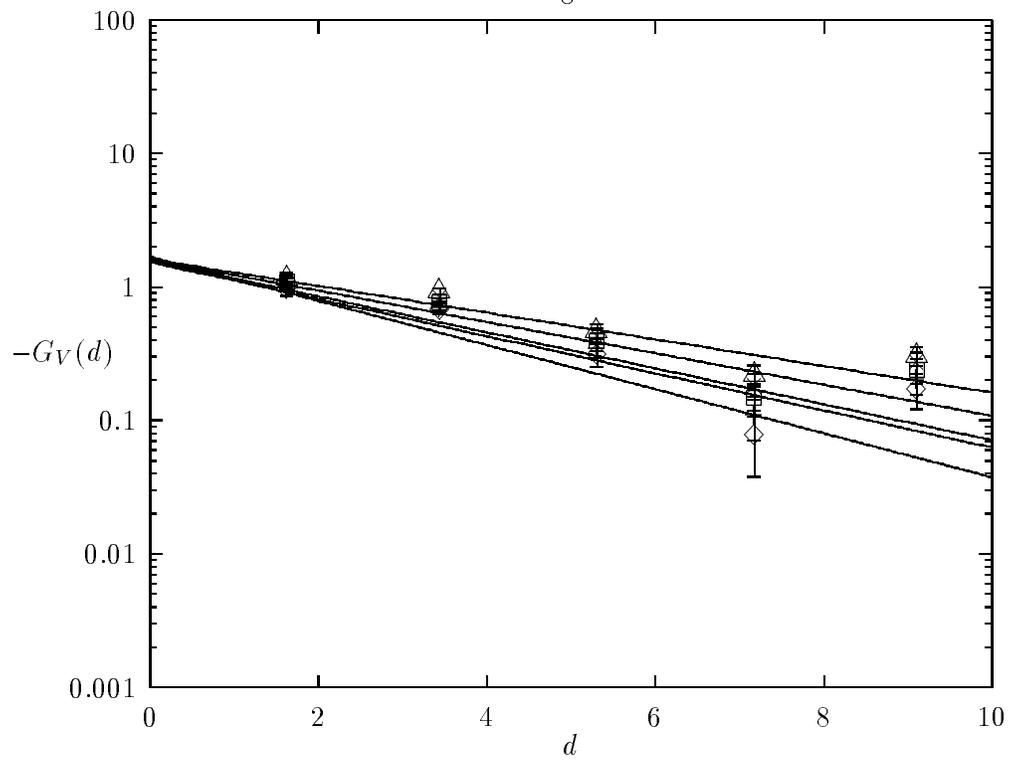

Fig. 2

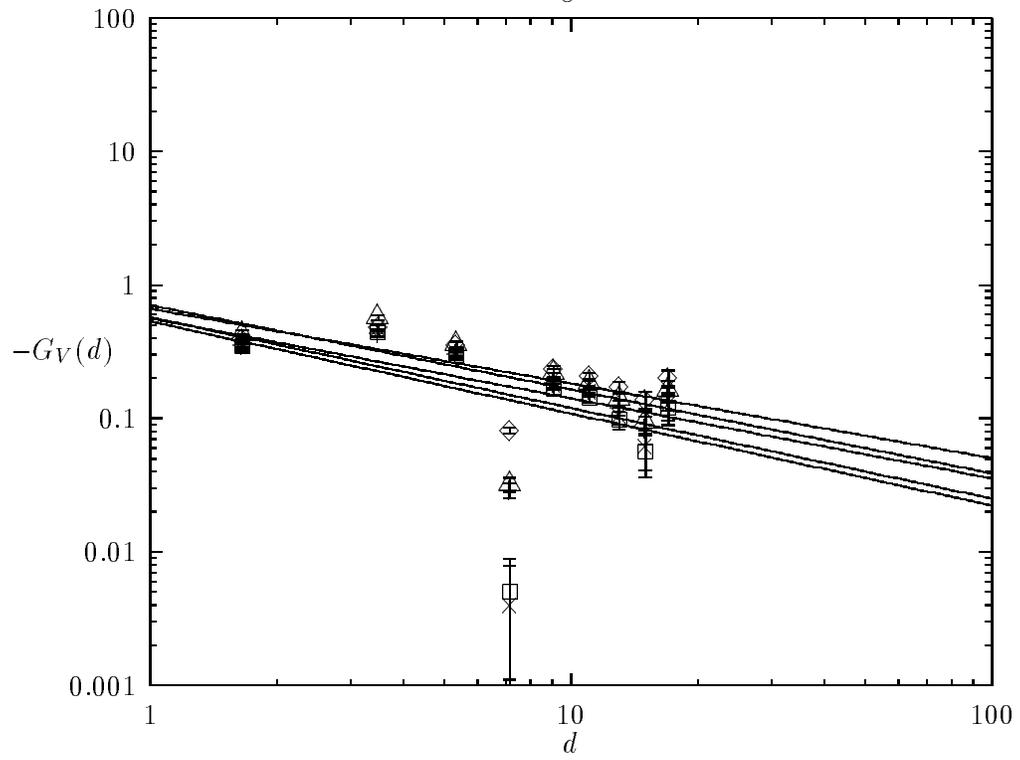

Fig. 3

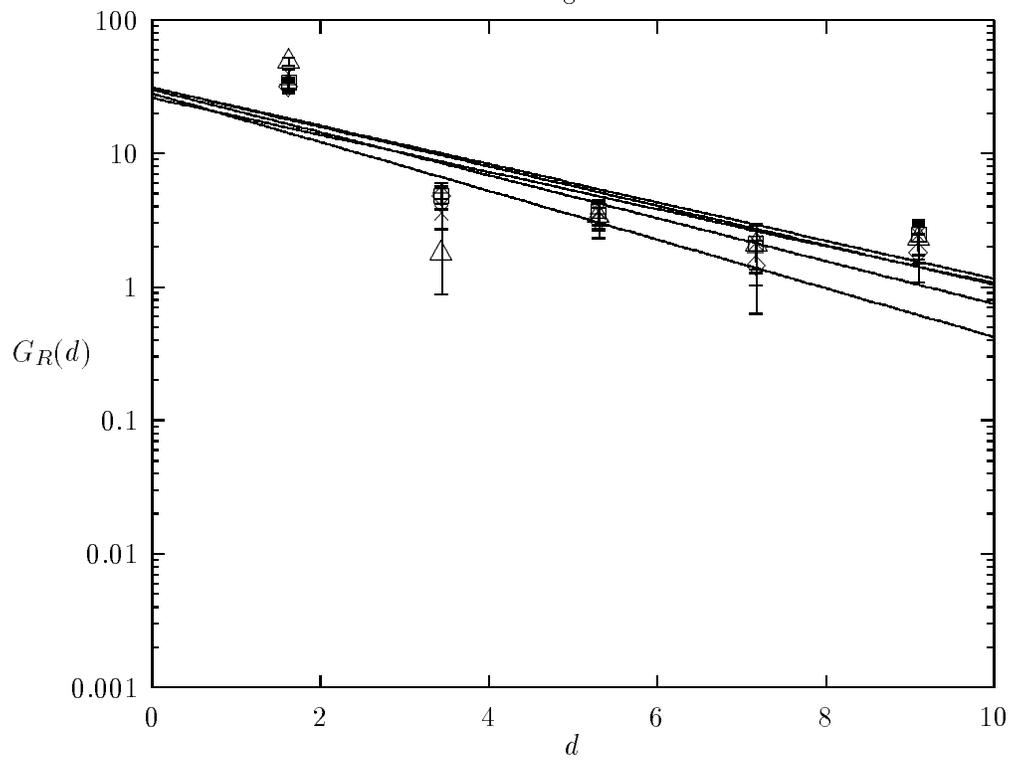

Fig. 4

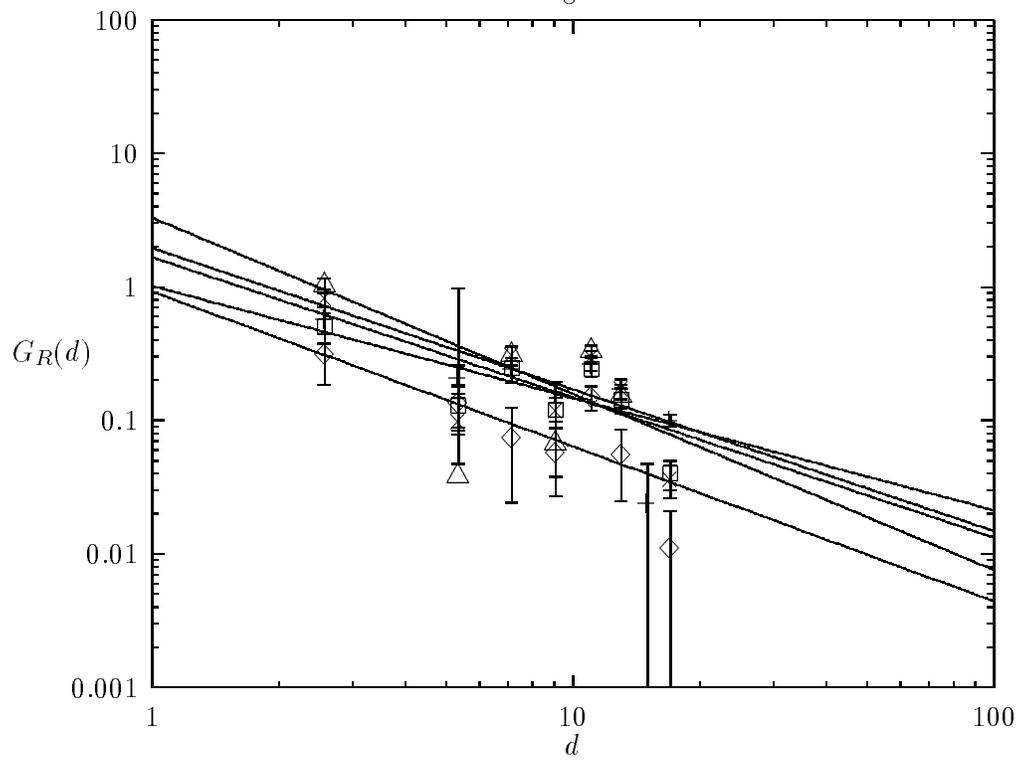

Fig. 5